\documentclass[pdflatex,sn-mathphys-num]{sn-jnl}

\usepackage{graphicx}%
\usepackage{multirow}%
\usepackage{amsmath,amssymb,amsfonts}%
\usepackage{amsthm}%
\usepackage{mathrsfs}%
\usepackage[title]{appendix}%
\usepackage{xcolor}%
\usepackage{textcomp}%
\usepackage{manyfoot}%
\usepackage{xcolor,pifont}
\newcommand{\cmark}{\ding{51}}%
\newcommand{\xmark}{\ding{55}}%
\usepackage{booktabs}%
\usepackage{algorithm}%
\usepackage{algorithmicx}%
\usepackage{algpseudocode}%
\usepackage{listings}%
\usepackage{tcolorbox}
\usepackage{enumitem}  
\usepackage{subcaption}
\usepackage{wrapfig}

\usepackage{booktabs}
\usepackage{tabularx}
\usepackage{array}
\usepackage{marvosym}

\newcolumntype{Y}{>{\hsize=1.5\hsize}X}  

\theoremstyle{thmstyleone}%
\theoremstyle{thmstyletwo}%

\theoremstyle{thmstylethree}%

\begin{document}

\title{AskDB: An LLM Agent for Natural Language Interaction with Relational Databases}

\author[1]{Xuan-Quang Phan\textsuperscript{$\dagger$}}
\author[1,3]{Tan-Ha Mai\textsuperscript{$\dagger$}}
\author[2]{Thai-Duy Dinh}
\author[2]{Minh-Thuan Nguyen}
\author[2]{Lam-Son L\^{e}\textsuperscript{\Letter}}

\affil[1]{IT Operations, Mantu Group, Vietnam}
\affil[2]{Faculty of Engineering, Vietnamese-German University, Vietnam}
\affil[3]{Computer Science and Information Engineering, National Taiwan University, Taiwan}

\affil[$\dagger$]{These authors contributed equally to this work}
\affil[\Letter]{Corresponding author: \texttt{lam-son.le@alumni.epfl.ch; son.ll@vgu.edu.vn}}


\abstract{Interacting with relational databases remains challenging for users across different expertise levels, particularly when composing complex analytical queries or performing administrative tasks. Existing systems typically address either natural language querying or narrow aspects of database administration, lacking a unified and intelligent interface for general-purpose database interaction.
We introduce AskDB, a large language model powered agent designed to bridge this gap by supporting both data analysis and administrative operations over SQL databases through natural language. Built on Gemini 2, AskDB integrates two key innovations: a dynamic schema-aware prompting mechanism that effectively incorporates database metadata, and a task decomposition framework that enables the agent to plan and execute multi-step actions. These capabilities allow AskDB to autonomously debug derived SQL, retrieve contextual information via real-time web search, and adaptively refine its responses.
We evaluate AskDB on a widely used Text-to-SQL benchmark and a curated set of DBA tasks, demonstrating strong performance in both analytical and administrative scenarios. Our results highlight the potential of AskDB as a unified and intelligent agent for relational database systems, offering an intuitive and accessible experience for end users.} 


\maketitle

\section{Introduction}
\label{sec:introduction}

Relational databases are the bedrock of modern information systems, yet accessing their full potential remains a significant challenge. For non-technical users, the complexity of structured query language (SQL) creates a barrier to direct data analysis \cite{Woods1973Lunar, Hendrix1978LIFER}. Concurrently, database administrators (DBAs) are often encumbered with repetitive administrative tasks (e.g., system monitoring, generating usage reports, user account management, backup verification), distracting them from strategic responsibilities like performance optimization and security hardening. This dual-sided friction leads to underutilized data assets and operational inefficiencies across organizations.

The advent of large language models (LLM) presents a transformative opportunity to bridge this human-database divide \cite{hong2024nextgen, shi2024employing}. Research has progressed along two primary, yet largely separate, tracks. Firstly, the field of natural language to SQL (NL2SQL) has evolved from early rule-based systems \cite{Copestake1990NLIDBSurvey} to highly capable transformer-based models that excel on complex benchmarks such as Spider \cite{Yu2018Spider, Liu2024NL2SQLSurveyLLM}. 
Secondly, AI has been applied to automate backend database administration, creating self-tuning systems \cite{Chaudhuri2007SelfTuningDB}, automatic index managers~\cite{Chaudhuri1997AutoAdmin} and data annotated~\cite{AI-annotated},
though these tools typically lack interactive, conversational capabilities.
\begin{figure}[htb]
    \vspace{-5pt}
    \centering
    \includegraphics[width=0.90\linewidth]{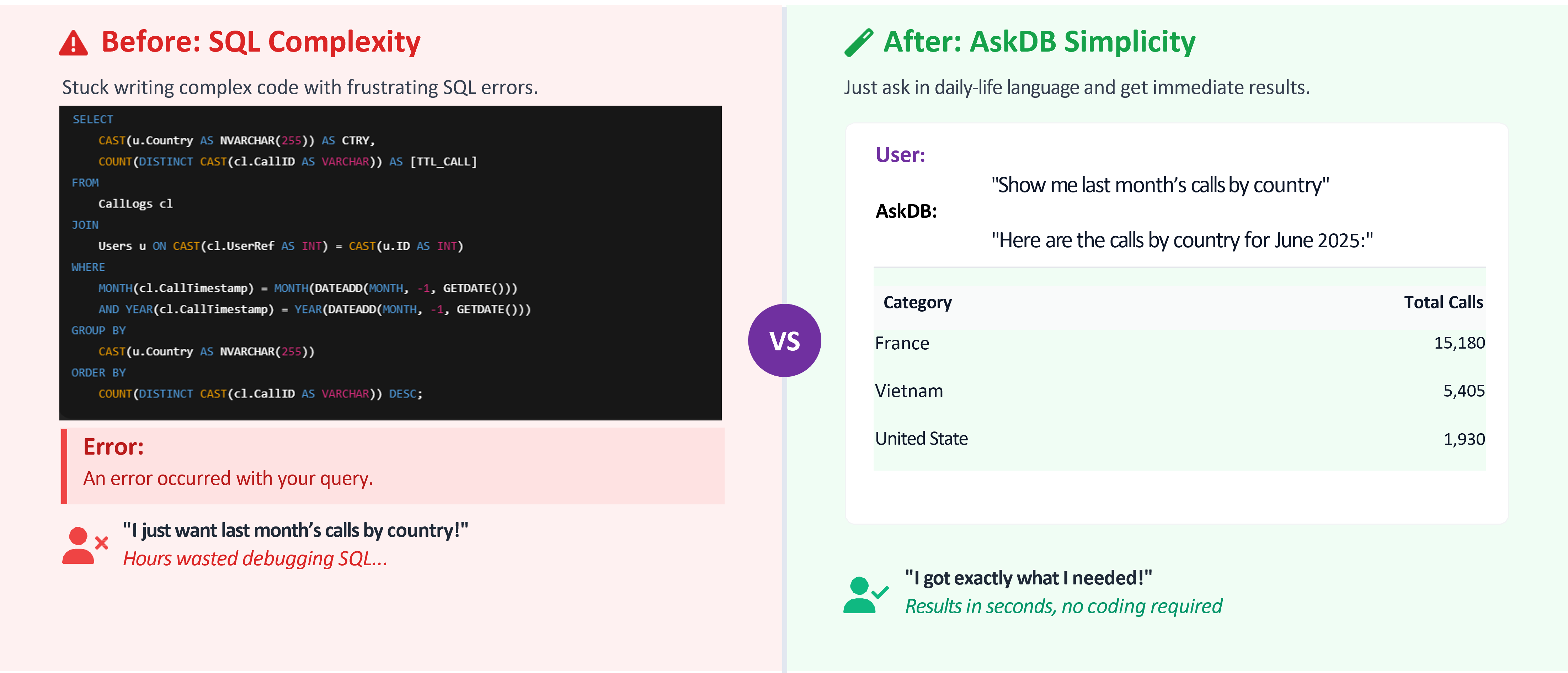}
    \caption{Illustration for the advantage of AskDB compared to traditional approach.}
    \label{fig:illustration-askdb}
    \vspace{-15pt}
\end{figure}

Despite these parallel advancements, a significant gap persists: the lack of a single highly conversational agent that can seamlessly handle both analytical queries from a business analyst and administrative commands from a DBA. Existing systems focus on either NL2SQL or DBA automation in isolation, failing to provide a single holistic interface for the full spectrum of database operations.

To address this gap, this paper introduces AskDB, a novel LLM-powered agent designed to unify data analysis and database administration through natural language. AskDB functions as an interactive co-pilot~\cite{GitHubCopilot2021}, leveraging a task decomposition framework and dynamic schema-aware prompting to interpret user intent, generate safe and accurate operations, and manage complex, multi-step tasks. It is built to assist a diverse range of users, from simplifying query generation to aiding in multifaceted administrative operations. As illustration in the Figure~\ref{fig:illustration-askdb}, our AskDB simplifies SQL-related tasks without the requirement for specialized SQL domain knowledge. Users can retrieve meaningful results effortlessly by expressing their queries naturally similar to human conversation via AskDB application. 
Our main contributions are threefold:
\begin{enumerate}[label=(\roman*)]
    \item \textbf{An overarching architecture} that integrates natural language data analysis and relational database administration capabilities into a single, cohesive system.
    \item \textbf{Novel methodologies for agentic database interaction}, including a dynamic schema-aware prompting mechanism for handling large schemas and an orchestrated framework that enables autonomous SQL debugging and internet-augmented problem-solving.
    \item \textbf{A comprehensive evaluation mechanism} demonstrating the proficiency of AskDB across both complex analytical query generation on a standard Text-to-SQL benchmark and a curated set of real-life database administration scenarios.
\end{enumerate}

The remainder of this paper is organized as follows: Section~\ref{sec:related_work} reviews related work in natural language interfaces to databases, LLM-based agents, and AI in database management. Section~\ref{sec:architecture_methodology} details the architecture and core methodologies of our proposed AskDB agent. Section~\ref{sec:experiment} describes the experimental setup, including datasets, evaluation metrics, and baselines. Section~\ref{sec:results} presents and discusses the results of our comprehensive evaluation. Finally, Section~\ref{sec:conclusion_future_works} concludes the paper, summarizing our findings and outlining directions for future research.
\section{Related Works}
\label{sec:related_work}

The development of AskDB is informed by two parallel, yet historically distinct, streams of research: facilitating natural language querying and automating database administration. We situate our work at the intersection of these fields, unified by the recent advancements in LLM-powered agentic systems.

\subsection{From NLIDBs to AI-Driven Administration}

The goal of querying databases via natural language has a rich history, beginning with early natural language interfaces to databases (NLIDBs) that relied on rule-based grammars and semantic parsing \cite{Woods1973Lunar, Hendrix1978LIFER, Copestake1990NLIDBSurvey}. While foundational, these systems were often brittle and struggled with linguistic variability. The field saw a significant leap with the advent of transformer-based architectures \cite{Brown2020GPT3, Raffel2020T5}, which now achieve remarkable performance \cite{lei2024spider2, hong2024nextgen, zhu2024enhanced}
 in complex query synthesis and schema linking on benchmarks like Spider. However, the primary focus of NL2SQL research remains on query generation.

Concurrently, a separate line of research applied AI to enhance database administration. This led to backend-oriented optimization systems including self-tuning databases \cite{Chaudhuri2007SelfTuningDB}, 
automatic index managers \cite{Chaudhuri1997AutoAdmin}, and cloud offerings such as oracle autonomous DB~\cite{OracleAutonomousDB2023}. These systems excel at automation but typically operate as blackboxes, lacking the fine-grained, conversational control necessary for interactive problem-solving or exploratory tasks.

These two research areas have evolved largely in isolation. AskDB bridges this divide by positioning itself not merely as a query generator or an automation tool, but as an \emph{interactive co-pilot}. It integrates the advanced query generation capabilities of modern NL2SQL systems with the ability to engage in a conversation about administrative tasks, which is a feature absent in traditional DBA automation.

\subsection{The Rise of LLM-Powered Agents}

The agentic capabilities of modern LLMs are the key enabler of the unified approach of AskDB. Foundational models such as Codex, GitHub Copilot, and Gemini \cite{Chen2021Codex, GitHubCopilot2021, Google2023GeminiTR} have demonstrated powerful code synthesis abilities. More importantly, recent frameworks have extended LLMs beyond simple generation, equipping them with tool-use capabilities and turning them into reasoning agents~\cite{liu2023agentbench} 
that can plan and execute actions.
AskDB extends this agentic paradigm to the specific domain of database interaction. Unlike generic code generators, it operates within a safety-governed loop, leveraging a curated set of tools for schema exploration, risk-aware execution, and dynamic context management. By maintaining conversation history and reasoning over it, AskDB can handle complex, multi-step tasks that require both SQL generation and administrative logic, a distinction that sets it apart from both legacy NLIDBs and autonomous DBA platforms. This aligns our work with the emerging paradigm of human-centered, LLM-powered co-pilots.

\section{Architecture and Methodology}
\label{sec:architecture_methodology}

\begin{figure}[htb]
    \centering
     \includegraphics[width=\linewidth]{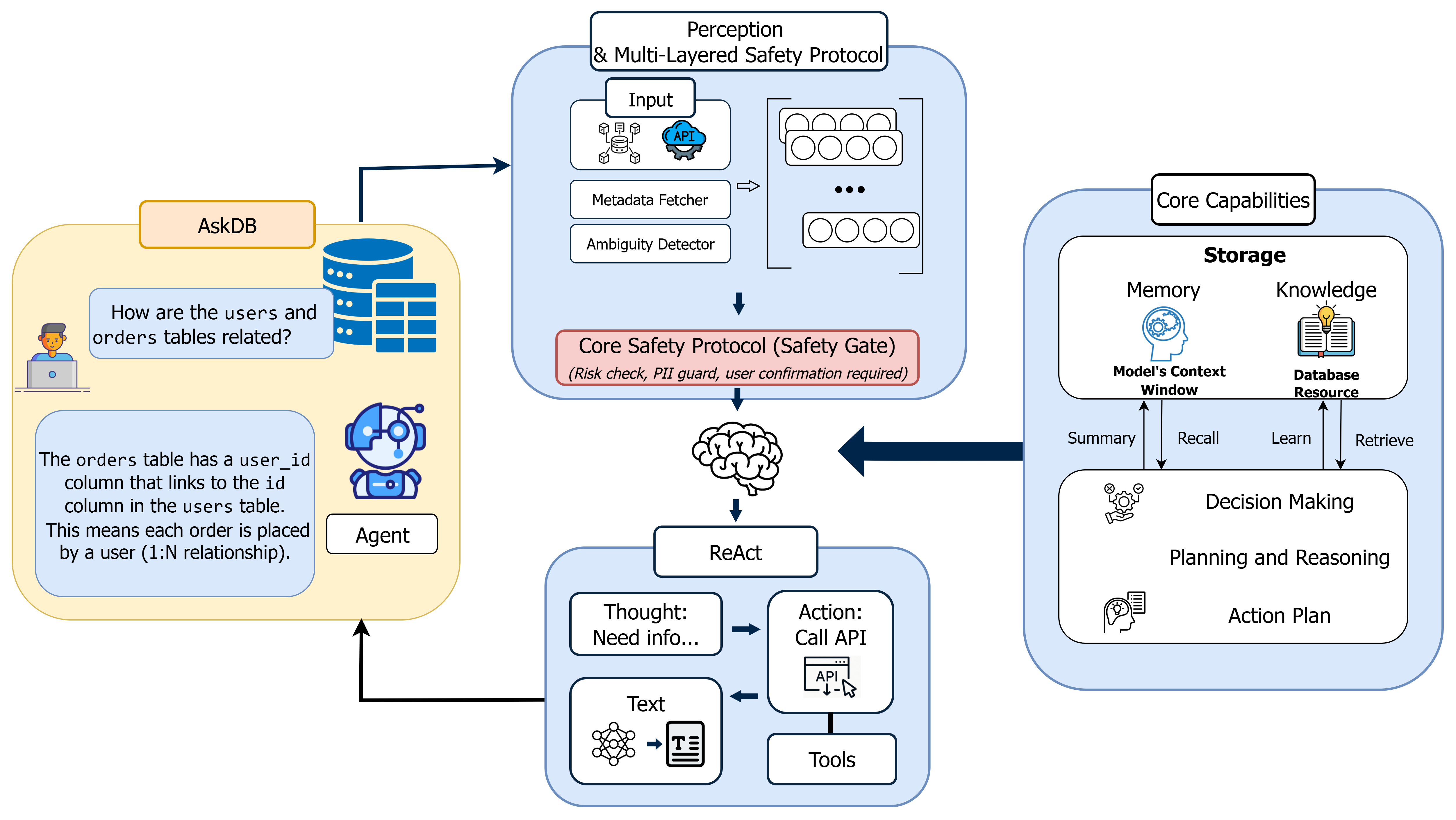}
     \caption{High-level conceptual architecture diagram of AskDB.}
     \label{fig:askdb_overall_architecture}
     \vspace{-10pt}
 \end{figure}

This section outlines the modular architecture and key mechanisms of AskDB, designed to function as an expert-level DBA and data analyst within a ReAct-driven workflow. Figure~\ref{fig:askdb_overall_architecture} presents the high-level architecture, which is further detailed below.

\subsection{The ReAct-Based Operational Framework}
\label{subsec:react_framework}

\begin{figure}[htbp]
    \centering
    \includegraphics[width=0.80\linewidth]{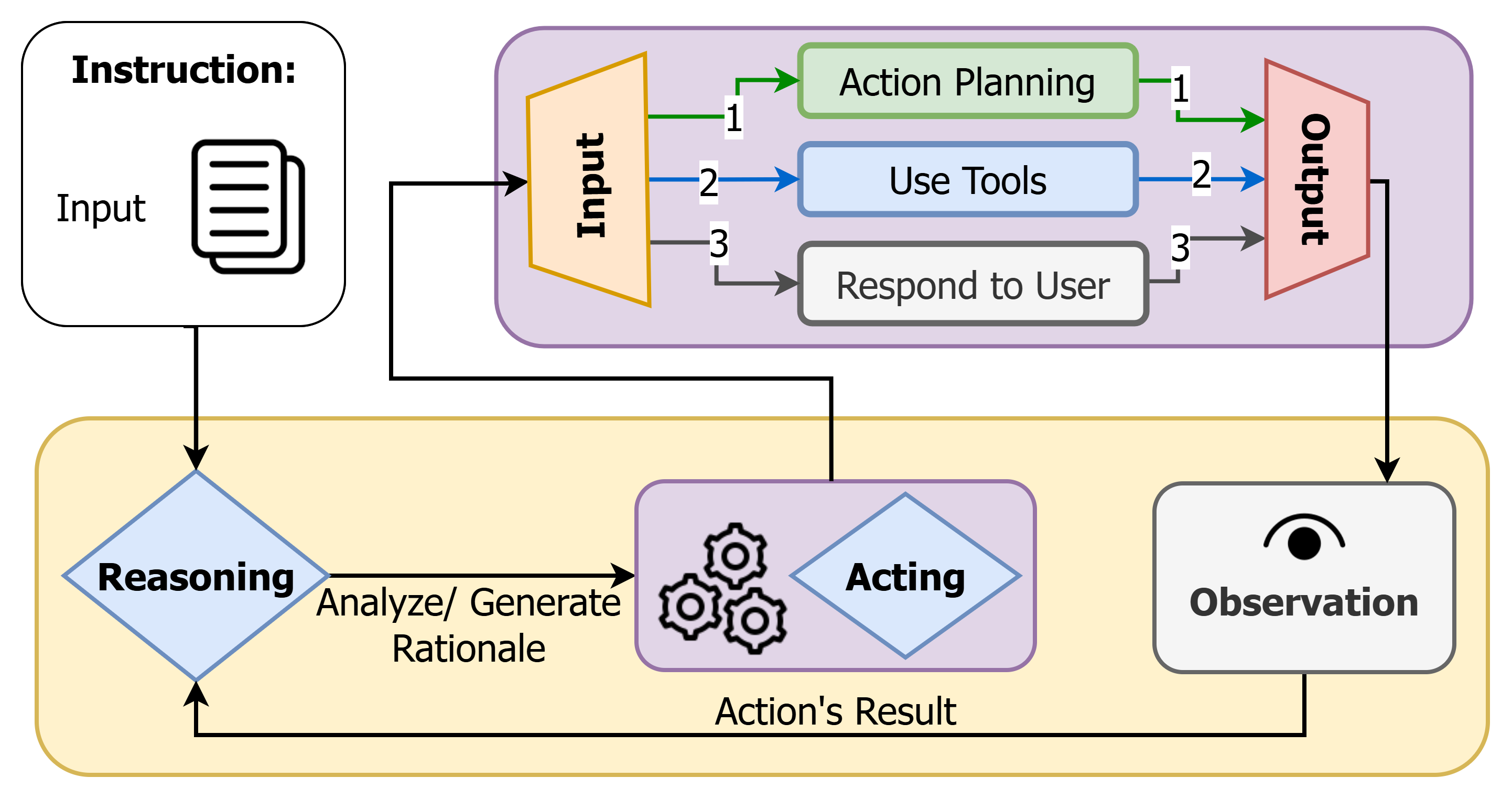}
    \caption{The ReAct-based operational framework of AskDB. The agent iteratively cycles through reason, act, and observe phases, driven by an LLM, to execute tasks.}
    \label{fig:react_framework}
    \vspace{-10pt}
\end{figure}

The heart of AskDB operation is the ReAct cognitive cycle, which governs every interaction and task. As illustrated in Figure~\ref{fig:react_framework}, this iterative, three-phase loop empowers the agent to handle complex problems that extend beyond simple one-turn query generation. The core cycle consists of:

\begin{itemize}
    \item \textbf{Reasoning:} The agent, powered by the LLM, first analyzes the current context, including the initial query of the user, the conversation history, and previous observations (if any). It then generates an explicit chain-of-thought rationale that explains \emph{why} a particular next action is necessary.
    \item \textbf{Acting:} Based on its reasoning, the agent makes a decision and executes a concrete action. This can range from invoking a specific tool from its arsenal (e.g., executing a query) to formulating a natural language response to ask the user for clarification.
    \item \textbf{Observation:} The result of the action can be an SQL query return, a database error message, the output of the tool, or the response to the user, which is captured as a new observation. This observation is then fed back into the start of the cycle, critically informing the next \texttt{Reasoning} phase.
\end{itemize}

In AskDB, the \texttt{Acting} phase of the ReAct cycle is a sophisticated decision-making step, not merely a code generation endpoint. The ability of the agent to plan and execute actions is significantly enhanced by its \emph{function calling capability}, a curated set of functions that extend its capabilities beyond the intrinsic knowledge of LLM. The primary tools include \texttt{execute\_query} for running \texttt{SELECT} statements, \texttt{execute\_non\_query} for state-modifying SQL commands, the \texttt{search\_tables\_by\_name} tool for schema exploration as described above, and \texttt{request\_for\_internet\_search} to retrieve external and information from the internet when needed.


The power of this tool-augmented, agentic approach is best demonstrated by the \emph{iterative SQL auto-debugging} loop of AskDB. When an \texttt{execute\_query} tool execution results in a database error, this error becomes an ``observation''. Instead of failing, the agent initiates a new ReAct cycle to resolve the issue

\begin{figure}[htbp]
    \centering
    \includegraphics[width=0.99\linewidth]{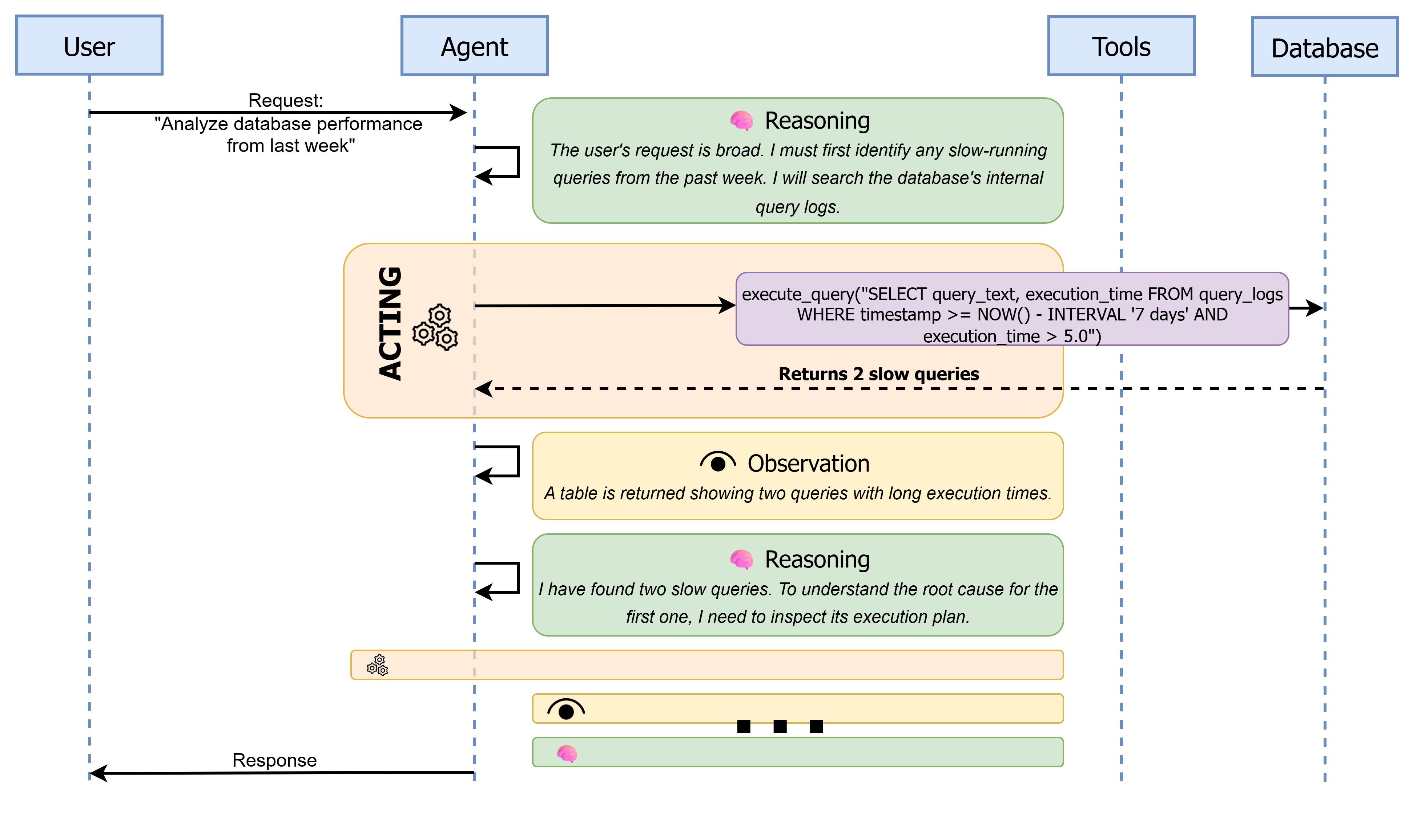}
    \caption{Step-by-step diagnostic reasoning process for identifying and resolving slow-running database queries. The system investigates recent logs, analyzes execution plans, identifies a full table scan on the orders table, and proposes indexing the ``order\_date'' column as a potential optimization.}
    \label{fig:react_seq_framework}
    \vspace{-10pt}
\end{figure}

This self-correction loop can continue for multiple iterations, allowing AskDB to autonomously resolve a wide range of common SQL errors such as incorrect column names, syntax mistakes, or faulty logic without requiring user intervention. This capability showcases a significant leap from simple generation to true problem-solving.
The power of this framework lies in its ability to transform a high-level, multi-step task into a coherent series of smaller, verifiable actions. For instance, faced with a request like \textit{``analyze database performance from last week''}, the agent does not attempt to answer in one shot. Instead, it initiates a ReAct trajectory as described in the Figure~\ref{fig:react_seq_framework}.

\subsection{Core Capabilities: Grounding, Planning, and Tool Use}
\label{subsec:core_capabilities}

The effectiveness of the ReAct operational framework is contingent upon the specialized capabilities of the agent  to perceive its environment, formulate sound plans, and execute actions. AskDB is endowed with a set of core capabilities designed specifically for the database domain, focusing on contextual grounding, agentic planning, and tool-augmented execution.

\paragraph{Contextual grounding via dynamic schema-aware prompting}
\label{sec:contextual_grounding}

A fundamental challenge for any database agent is \textit{grounding}, which connects the natural language query of the user  to the specific, and often complex structure of a target database. LLMs have no intrinsic knowledge of a given schema, and it is infeasible to provide the entire schema for a large database due to limitations of the context window and the risk of distracting the model with irrelevant information.

\begin{figure}[htbp]
    \centering
    \includegraphics[width=0.95\linewidth]{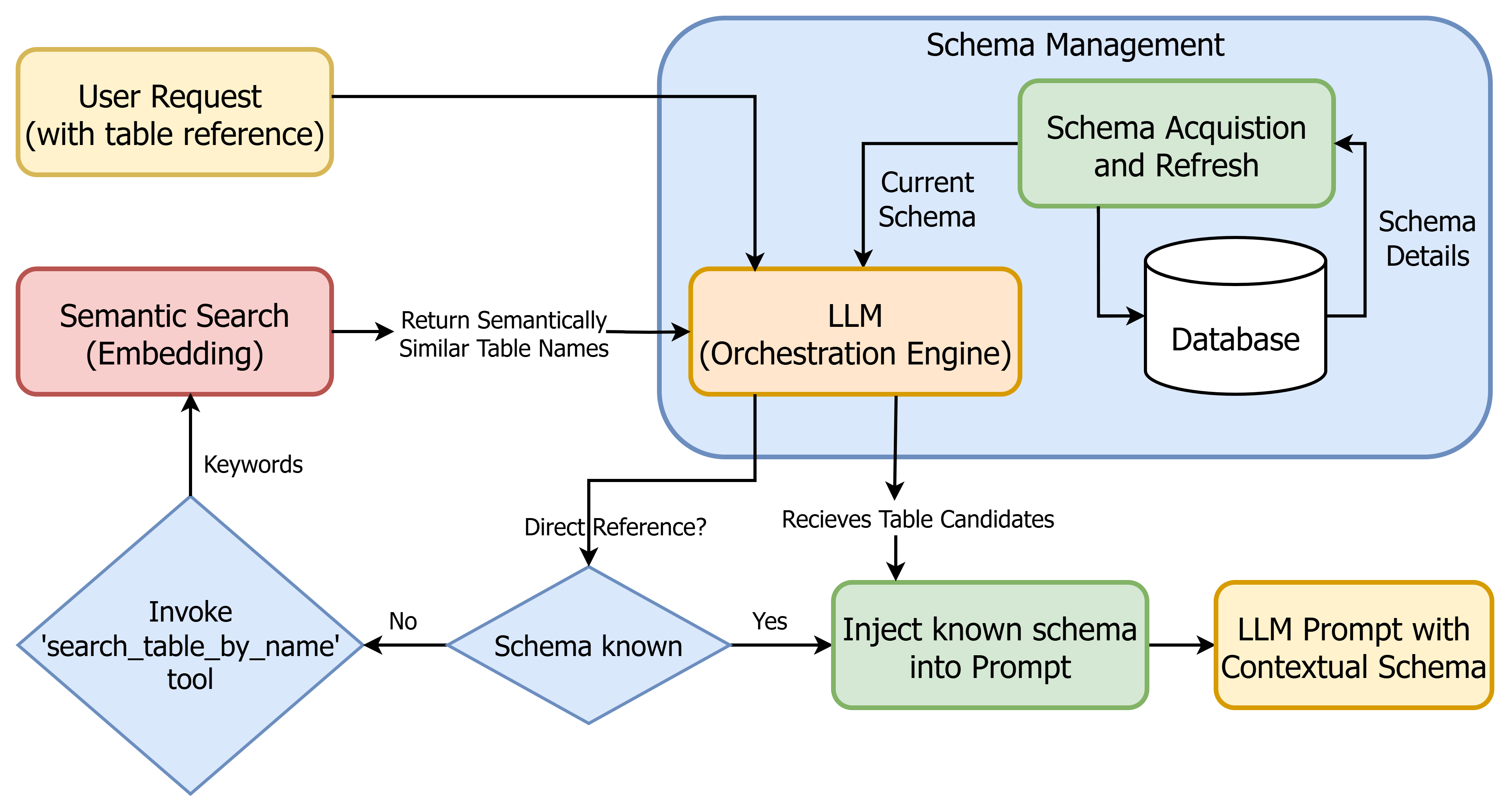}
    \caption{The dynamic schema-aware prompting mechanism. The agent uses semantic search to identify relevant tables and injects only their scoped schema information into the prompt, grounding the LLM in the current database context. This RAG-like approach mitigates challenges with large schemas and improves query accuracy.}
    \label{fig:schema_integration}
\end{figure}

To overcome this, AskDB employs a \emph{dynamic schema-aware prompting} mechanism, a retrieval-augmented generation (RAG)-like pipeline illustrated in Figure~\ref{fig:schema_integration} that leverages tool usage. This process ensures the LLM receives only the most relevant schema information for the task at hand:

\begin{itemize}
    \item \textbf{Entity recognition:} The agent first performs an initial pass on the query of the user to identify potential table or column name references.
    \item \textbf{Disambiguation via semantic search:} If references are ambiguous (e.g., user asks about ``customer data'' which could exist in ``customers'', ``users'', or ``clients'' tables) or if exploration is needed, the agent invokes the \texttt{search\_tables\_by\_name} tool. This tool leverages a pre-computed, in-memory vector index of all table names, created using the \texttt{all-MiniLM-L6-v2} sentence-transformer model. It performs a cosine similarity search to return a ranked list \cite{li2024metasql, sun2023dbcopilot} of the most semantically relevant table candidates.
    \item \textbf{Scoped context injection:} The schemas for only these top-ranked candidate tables are retrieved. Critically, this information is formatted into a structured markdown table that explicitly details column names, data types, and a dedicated \texttt{Constraint} column. This column clearly defines primary and foreign key relationships (e.g., \texttt{FOREIGN KEY (user\_id) REFERENCES users(id)}), which is vital for guiding the LLM to generate correct \texttt{JOIN} conditions.
    \item \textbf{Prompt assembly:} The final prompt sent to the LLM is a rich composite of the original user query, the relevant conversational history, and this precisely scoped, structured schema context.
\end{itemize}

This dynamic grounding mechanism not only makes interaction with large, complex schemas tractable but also significantly improves the accuracy of the generated SQL queries by focusing the attention of LLM on the pertinent parts of the database structure.

\subsection{The Multi-Layered Safety Protocol}
\label{subsec:safety_protocol}

The deployment of an autonomous agent to interact with critical data systems requires a foundational commitment to safety and reliability. To this end, AskDB integrates a \textit{multi-layered safety protocol}, which combines proactive user confirmation with automated, context-aware guardrails to mitigate risk before any potentially harmful operation is executed.

\paragraph{Proactive risk classification and confirmation}

The first layer of defense is a proactive risk assessment process. Every user request undergoes an initial analysis where the agent classifies the intent and potential impact as either \texttt{Low-Risk} or \texttt{High-Risk}. A request is deemed \texttt{High-Risk} if it involves any of the following:\textit{ (i)} Any state-modifying commands (e.g., \texttt{UPDATE}, \texttt{DELETE}, \texttt{DROP}); \textit{(ii)} Queries targeting tables identified as potentially large or containing sensitive information; \textit{(iii)} Ambiguous administrative requests that lack specific parameters (e.g., ``clean up old logs'').

\begin{figure}[htbp]
    \centering
    \includegraphics[width=0.95\linewidth]{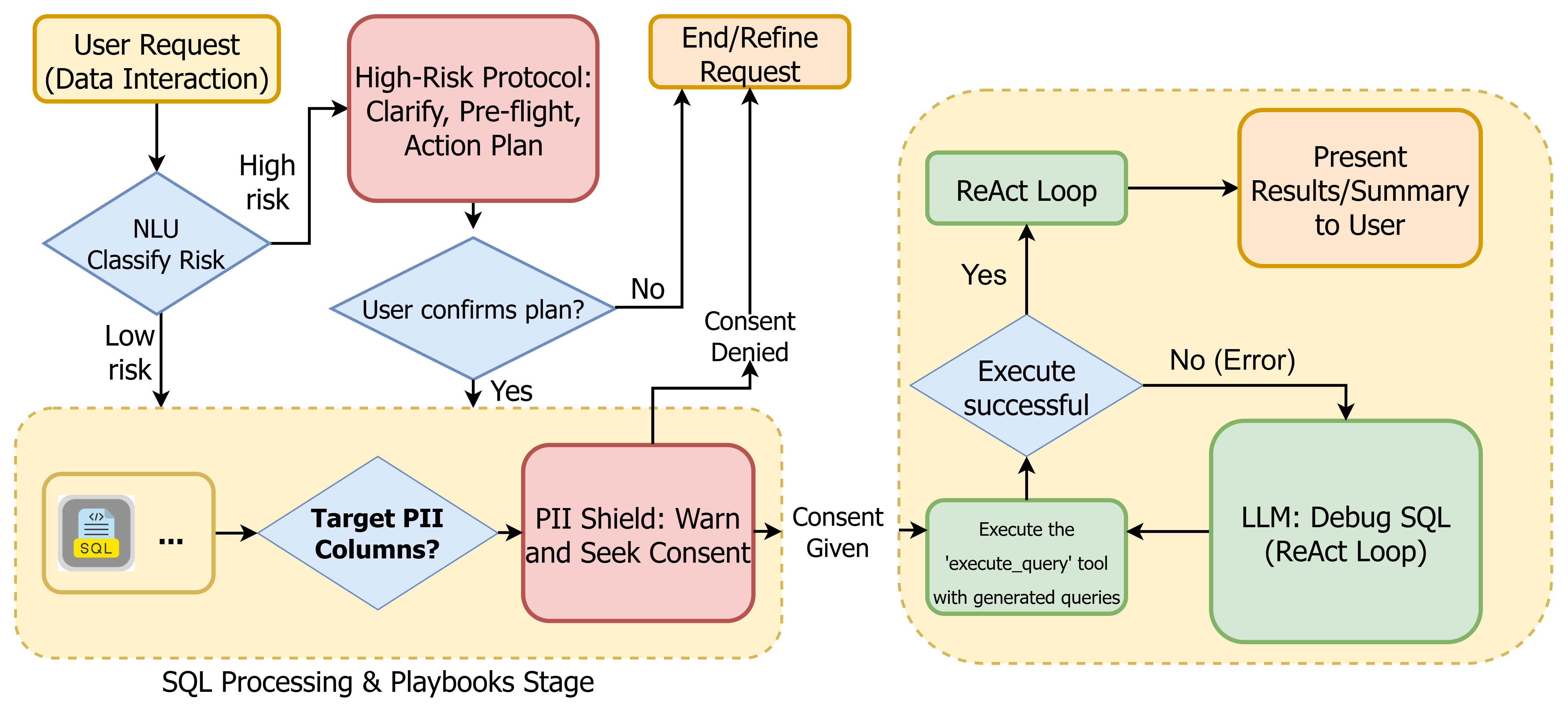}
    \caption{The \emph{multi-layered safety protocol} part from [\ref{fig:askdb_overall_architecture}] for data interaction. All requests undergo risk classification, with high-risk tasks requiring a user-confirmed Action Plan. Automated playbooks provide an additional layer of protection before safe execution.}
    \label{fig:safety_protocol_data}
\end{figure}

For any task classified as \texttt{High-Risk}, AskDB enters a mandatory confirmation loop. The agent must halt direct execution, formulate a clear, multi-step action plan, and present this plan to the user with explicit warnings about its potential consequences. Only after receiving explicit confirmation (e.g., the user typing ``yes'' or ``proceed") will the agent execute the plan. This human-in-the-loop design ensures user oversight for all critical operations.

\paragraph{Automated guardrail playbooks}

The second layer consists of specialized, automated guardrail playbooks that are triggered by specific patterns during the reasoning process of the agent. These playbooks enforce best practices and prevent common pitfalls without requiring constant user intervention for routine issues.

\begin{enumerate}[label=(\roman*)]
    \item \textbf{Sensitive data protection (PII shield):} Instead of pre-defined keyword filtering, the agent uses instructed LLM reasoning to semantically detect personally identifiable information (PII), including subtle fields like \texttt{mother\_maiden\_name} or \texttt{user\_bio}. When such data is requested, the PII shield triggers the \textbf{Layer 1} \texttt{High-Risk} confirmation, requiring explicit user approval before access.

    \item \textbf{Performance and efficiency guardrails:} To promote efficient database usage, the agent employs a \textbf{\texttt{SELECT *} Interception} playbook. When a user enters a broad query like \texttt{SELECT * FROM large\_table}, the agent will silently intercept the request, explain the performance implications, and ask the user to specify the exact columns they need before proceeding.

    \item \textbf{System integrity guardrails:} For the most dangerous administrative tasks, AskDB utilizes a \textbf{destructive operation playbook}. This enforces a ``double confirmation'' procedure for commands like \texttt{TRUNCATE} or \texttt{DROP}. The agent will not only present an action plan but will also ask for a second, explicit confirmation, ensuring the user is fully aware of the irreversible nature of the operation.
\end{enumerate}
By combining proactive, user-driven confirmation with automated, intelligent guardrails, the multi-layered protocol of AskDB creates a robust safety net. This allows the agent to be both highly capable and trustworthy, a crucial prerequisite for its role as a database co-pilot.
\section{Experiment}
\label{sec:experiment}
\vspace{-5pt}
\subsection{Technical Stack}
\label{subsec:experimental_setup}
The application was developed using C\# and XAML, leveraging the .NET 8 in combination with WinUI 3 for the user interface and for database access. For natural language processing, the system integrates the Gemini 2.0 Flash and Gemini 2.5 Flash LLMs. The backend supports multiple database engines, including SQL Server, PostgreSQL, MySQL, SQLite, and MariaDB, allowing flexible deployment and cross-database compatibility.

\subsection{Evaluation Methodology}
\label{sec:eval_method}
\paragraph{Evaluation Metric}
Traditional single-turn accuracy is insufficient to measure the multi-step, agentic process of AskDB. The primary metric is \emph{Interaction Efficiency}, defined by the number of interaction turns required to complete a user request \cite{walker1997paradise}. Fewer turns indicate greater efficiency and autonomy, as the agent can resolve complex tasks with minimal user input, thus reducing user effort and improving overall experience.

\paragraph{Benchmarks}
The evaluation of Text-to-SQL capabilities uses the widely recognized Spider benchmark, including both the original Spider 1.0 dataset \cite{Yu2018Spider} and the more challenging Spider 2.0 set \cite{lei2024spider2}. Query correctness is determined by \emph{Execution Accuracy} \cite{Yu2018Spider}, which requires the executed SQL to return a result set equivalent to the ground-truth answer. This metric ensures both syntactic validity and semantic correctness. As AskDB is designed for interactive use, a full evaluation on the Spider benchmarks is impractical, so we selected a representative subset for evaluation.

\paragraph{Zero-Shot Autonomous Execution}
To evaluate autonomy, we utilize the protocol simulates a non-technical user. 
\begin{wrapfigure}{r}{0.39\textwidth}
    \vspace{-15pt}
    \includegraphics[width=0.38\textwidth]{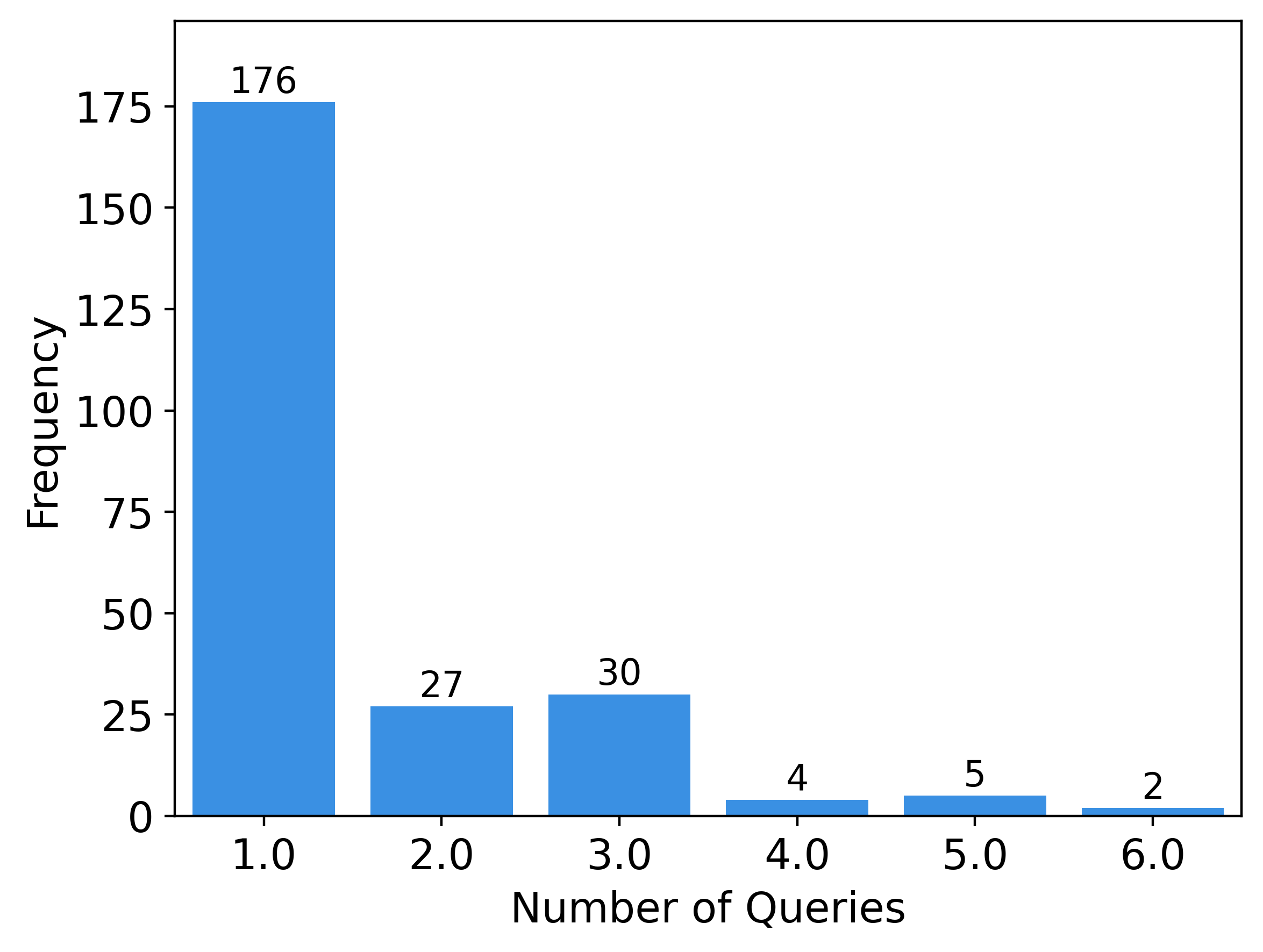}
    \caption{Distribution of interaction attempts required to resolve queries on Spider 1.0.} 
    \label{fig:spider1_attempts}
    \vspace{-15pt}
\end{wrapfigure}
Each evaluation starts with only the benchmark's natural language question, without schema hints or query refinements. 
When confirmation is required, such as schema exploration or high-risk actions, the simulated user gives standardized approval (e.g., ``yes, proceed''), allowing the agent to gather context and resolve ambiguities independently. Any necessary external knowledge (e.g., domain-specific specification) is preloaded and does not count as an interaction turn. This approach ensures the results reflect the ability of AskDB to operate with minimal human intervention.

\section{Results and Discussion}
\label{sec:results}

\subsection{AskDB agent implementation}

AskDB has been implemented as an Windows 11 application with the technical stack mentioned in \ref{subsec:experimental_setup}. Below are some screenshots from the application. In addition, AskDB gained \emph{over 200 active users} since the first launch in June 2025.

\begin{figure}[htb]
    \centering
    \includegraphics[width=0.95\linewidth]{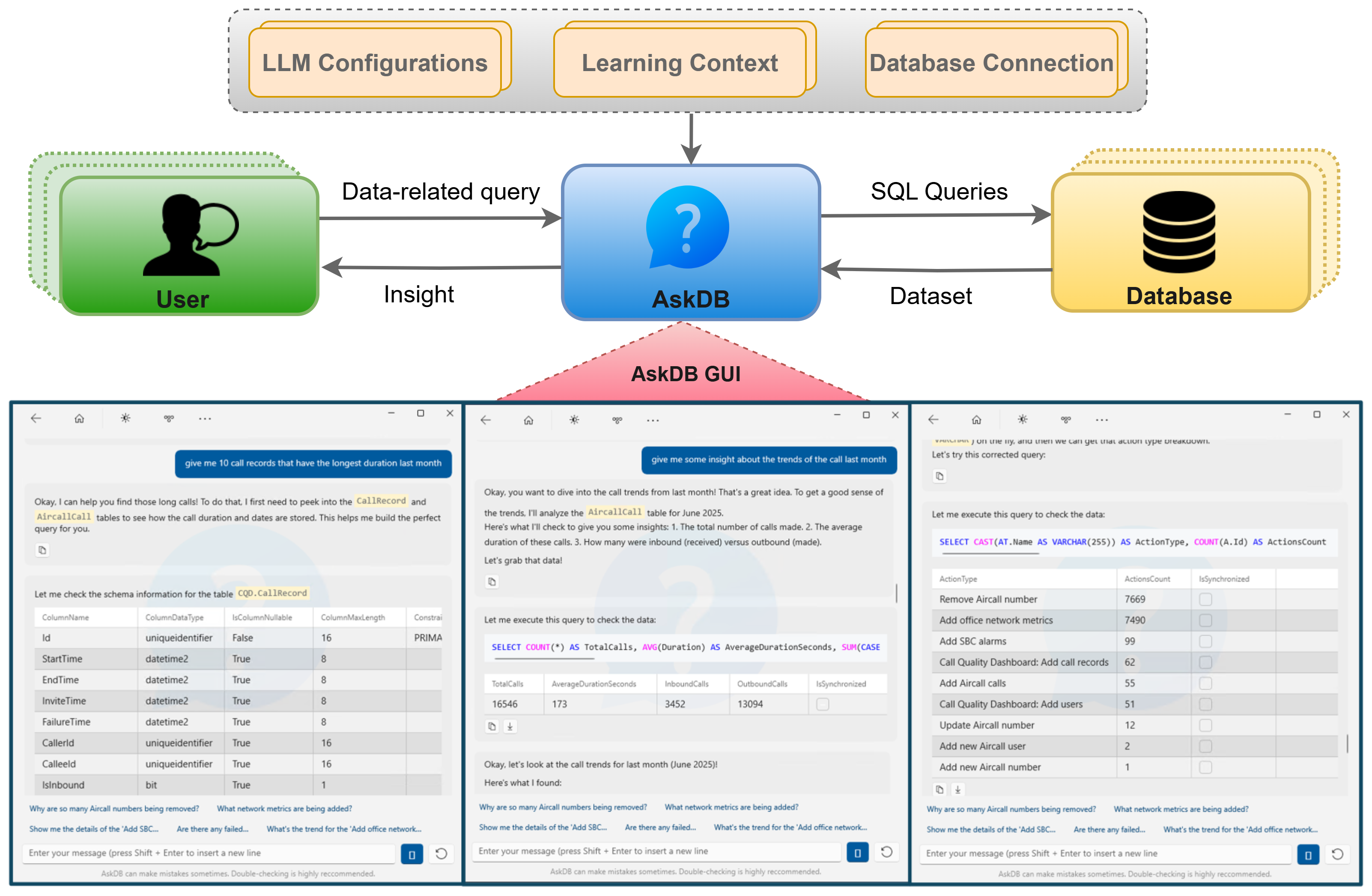}
    \caption{Real-world application of AskDB in enterprise data systems.}
    \label{fig:app_demo}
\end{figure}


\subsection{Performance on Spider 1.0 Benchmark}
For the Spider 1.0 evaluation, we selected 8 distinct databases, testing the agent on a total of 98 questions. This selection was designed to cover a broad spectrum of complexity, spanning across different domain, from easy to extra-hard questions and encompassing multiple database tasks such as filtering, aggregation, and multi-table joins. AskDB achieved an overall execution accuracy of 89.8\% on this diverse subset of Spider 1.0. Despite the agent is not provided with the database schema except table names, its performance is highly competitive, even surpassing several LLM-based Text-to-SQL benchmarks on Spider 1.0 as shown in Table \ref{tab:model_comparison}.

\begin{table*}[h!]
\vspace{-10pt}
\centering
\caption{Performance of AskDB on Spider datasets}
\label{tab:spider1_results}
\scalebox{0.9}{
\begin{tabularx}{1\textwidth}{@{} l X X X X X X @{}}
\toprule
Dataset & Hardness level & Total question & Correct answers & Average retrieval & Accuracy\\
\midrule
 & Easy & 16 & 16 & 1.33 & 100\% \\
& Medium & 31 & 31 & 1.41 & 100\%\\
Spider 1.0 & Hard & 19 & 15.33 & 1.57 & 80.7\%\\
& Extra Hard & 32 & 25.67 & 1.47 & 80.2\% \\
& All & 98 & 88 & 1.45 & 89.8\% \\
\midrule
 & Easy & 9 & 5.67 & 1.44 & 62.96\%\\
Spider 2.0 & Medium & 52 & 19.33 & 1.14 & 37.18\%\\
& Hard & 51 & 15.67 & 1.52 & 30.72\%\\
& All & 112 & 34.33 & 1.34 & 36.31\%\\
\bottomrule
\end{tabularx}}
\vspace{-7pt}
\end{table*}
Furthermore, the average retrieval value of 1.45 indicates high interaction efficiency, meaning AskDB requires minimal conversational turns to resolve queries. This low interaction count demonstrates the agent's autonomy, as it can resolve complex tasks with reduced user input, thereby enhancing the overall user experience.On this subset, AskDB achieved an execution accuracy of 36.31\%, requiring an average of only 1.34 inquiries per question. This performance is particularly noteworthy, rivaling methods that deploy far more complex LLM models and demonstrating strong generalization capabilities across distinct datasets.

\subsection{Performance on Spider 2.0 Benchmark}
Spider 2.0 is a more challenging and realistic benchmark than its predecessor, designed to reflect real-world enterprise SQL workflows. Compared to Spider 1.0, it features more complex databases with an increased number of columns and tables, support for various database dialects, and more sophisticated query requirements within an agentic task setting.
We evaluated the performance of AskDB on Spider 2.0-lite benchmark, focusing on the version that supports SQLite databases. For this evaluation, a subset of 90 questions was selected from 25 different databases, which included questions requiring external knowledge. 
\begin{table*}[h]
\centering
\caption{Performance and cost comparison between various LLM-based agents}
\label{tab:model_comparison}
\scalebox{0.85}{
\begin{tabular}{@{}lllccc@{}}
\toprule
Agent & Dataset & Model & Reasoning & Cost\footnotemark[1] & Accuracy\footnotemark[4] \\
\midrule
C3~\cite{dong2023c3} & &GPT-3.5-Turbo+Zero-Shot & \xmark & \$1/\$4 & 82.3\%\\
DAIL-SQL~\cite{gao2023texttosql} & & GPT-4+Self-Consistency & \xmark & \$30/\$60 & \underline{86.6\%} \\
DAIL-SQL~\cite{gao2023texttosql} & Spider 1.0 & GPT-4 & \xmark & \$30/\$60 & 86.2\% \\
DIN-SQL~\cite{pourreza2023dinsql} & & Codex & \xmark & Deprecated & 78.2\%\\
AskDB (\textbf{\textit{ours}}) & & Gemini 2.0 Flash & \xmark & \$0.1/\$0.4\footnotemark[2] & \textbf{89.8\%} \\
\midrule
ReFoRCE~\cite{deng2025reforce} & &  o3 & \cmark & \$2/\$8 & \textbf{37.8\%} \\
RSL-SQL~\cite{cao2024rslsql} & & o3 & \cmark & \$2/\$8 & 33.1\%  \\
Spider-Agent & Spider 2.0 & GPT-4o-2024-11-20 & \xmark & \$2.5/\$10 & 13.16\% \\
LinkAlign~\cite{wang2025linkalign} & & DeepSeek-R1 & \cmark & \$0.25/\$2.19 & 33.09\%\\
AskDB (\textbf{\textit{ours}}) & & Gemini 2.5 Flash & \cmark & \$0.15/\$0.60\footnotemark[2] & \underline{36.31\%} \\
\bottomrule
\end{tabular}}
\end{table*}
\footnotetext[1]{Costs are listed as ``input/output price'' per 1 million tokens. API prices retrieved on July 15, 2025.}
\footnotetext[2]{As of July 15, 2025, Google also offers a free tier for these models: up to 200 requests per day for Gemini 2.0 Flash and 250 requests per day for Gemini 2.5 Flash.}
\footnotetext[4]{Testing results of different model retrieved at \url{https://spider2-sql.github.io/} on July 15, 2025.}

A key advantage of our approach is its cost-efficiency of the product compared to current LLMs benchmarked on these datasets. This efficiency is archieved by utilizing lightweight models such as Google's Gemini 2.0 Flash and Gemini 2.5 Flash. These model were selected not only for their great free tier and inexpensive pricing offering, but they have minimal intrinsic reasoning capabilities. These lightweight architecture allows the system to generate an answer promptly, therefore, enhance the user experience when compared to solutions built on larger, more computationally intensive LLMs. By leveraging cost-efficiency models like Gemini 2.0/2.5 Flash, AskDB achieves competitive performance without the significant computational and financial overhead of larger models like GPT-4. This makes our solution more practical for real-world deployment, especially in scenarios requiring high throughput.

\section{Conclusion and Future Works}
\label{sec:conclusion_future_works}
AskDB represents a significant advancement in bridging the gap between natural language and relational databases. Our primary contribution is a \textit{ReAct-based cognitive framework} that unifies NL2SQL and DBA tasks within a single, autonomous agent. This approach enables AskDB to handle complex, multi-step requests, moving beyond simple query generation. Empirical evaluations validate this capability, with the agent achieving a strong 89.8\% execution accuracy on a representative subset of the Spider~1.0 benchmark, demonstrating high interaction efficiency with minimal user intervention.

However, our work also highlights clear limitations and opportunities for future enhancement. The agent's 36.31\% accuracy on the more challenging Spider~2.0 benchmark indicates a current weakness in handling deeply nested queries and highly complex semantic logic. Architecturally, the system's present implementation is dependent on Google's Gemini models, and its performance has only been validated on benchmark subsets. Future work will focus on three key areas: (1) Enhancing the agent's core reasoning capabilities to master more complex query scenarios; (2) Expanding the experimental scope to full benchmark datasets to ensure robust evaluation; and (3) Decoupling the agent from a single LLM provider by developing a provider-agnostic architecture, which will also pave the way for self-hosting options. Addressing these limitations will be crucial in developing AskDB into a truly versatile and practical co-pilot for database interaction.

\bibliography{references}

\end{document}